\documentclass[12pt]{article}
\usepackage{amssymb,dsfont,amsmath,amsfonts,hyperref}
\usepackage{slashed}
\usepackage{color}

\def\be{\begin{equation}}
\def\ee{\end{equation}}

\def\a{\alpha}
\def\b{\beta}

\begin{document}

\title{A Higher-Spin Chern--Simons Theory of Anyons\footnote{Based on talk given by M. V. at \textit{``Supersymmetries and Quantum Symmetries''}, Dubna,
July 29 - August 3, 2013 with proceedings to appear in \textit{Physics of Particles and Nuclei}.}}

\author{
N. Boulanger$^1$,
P.  Sundell${}^2$ \& M.  Valenzuela${}^{1,3}$ \\[10pt]
{ \small
${}^1$\textit{Service de M\'ecanique et Gravitation,
Universit\'e de Mons, Belgique.} } \\
{ \small ${}^2$\textit{Departamento de Ciencias F\'isicas, Universidad Andres Bello, Santiago, Chile.}}\\
{ \small ${}^3$\textit{Instituto de Ciencias F\'isicas y Matem\'aticas, Universidad Austral de Chile, Valdivia, Chile.}}}

\date{}

\maketitle

\begin{abstract}
We propose Chern--Simons models of fractional-spin fields interacting with
ordinary tensorial higher-spin fields and internal color gauge fields.
For integer and half-integer values of the fractional spins, the model reduces  
to finite sets of fields modulo infinite-dimensional ideals.
We present the model on-shell using Fock-space representations of the 
underlying deformed-oscillator algebra.
\end{abstract}

\vspace{5mm}

Quantum mechanics in $2+1$ dimensions admits anyons: fractional-spin 
particles \cite{Leinaas:1977fm} with correlated generalized statistics properties
governed by the braid group \cite{Forte:1990hd}.
Anyons arise in a number of systems; for example, as 
non-relativistic charged vortices \cite{Wilczek:1982wy}, 
relativistic Hopf-interacting massive particles in matter-coupled 
Chern--Simons theories \cite{Polyakov:1988md,Semenoff:1988jr} 
and vertex operators in two-dimensional conformal field theories \cite{Jatkar:1990re}.
In this letter, we present topological models of Chern--Simons 
type that describe fractional-spin gauge fields coupled to higher-spin gravities 
(HSGRA) and internal gauge fields, which we will refer to as 
fractional-spin gravities.
Diffeomorphism invariance as well as higher-spin symmetries 
are indeed natural in theories of anyons, 
essentially due to the topological character of the local rotations 
and translations underlying the generalized spin-statistics relations \cite{Forte:1990hd} 
and 
the fact that local constructs built out of fractional-spin fields 
decompose under the Lorentz algebra into infinite towers of higher-spin 
Lorentz tensors and tensor-spinors.

Our construction stands on the observation that the Prokushkin--Vasiliev
system \cite{Prokushkin:1998bq}, which provides the only known fully non-linear 
description of three-dimensional matter-coupled HSGRA, 
admits several inequivalent embeddings of the Lorentz algebra 
into its higher-spin algebra \cite{wipFS1}, 
besides the standard embedding leading to tensor-spinorial HSGRA.
The Prokushkin--Vasiliev system consists of a connection one-form 
$\widehat A$ and matter zero-form $\widehat B$ living on a base manifold given 
locally by the direct product of a commutative spacetime ${\cal M}$ 
and non-commutative twistor space ${\cal Z}$ with a closed and central two-form $\widehat J\,$.
These master fields are valued in associative algebras consisting of functions 
on a fiber manifold ${\cal Y}\times {\cal I}\,$, 
the product of an additional twistor space ${\cal Y}$ 
and an internal manifold ${\cal I}$ whose coordinates 
generate a matrix algebra; for further details, see \cite{wipFS1}.
The Prokushkin--Vasiliev field equations, \emph{viz.} 
$\widehat {\rm d}\widehat A+\widehat A^2+\widehat J \widehat B=0$ and 
$\widehat {\rm d}\widehat B+[\widehat A,\widehat B]=0$, 
state that $\widehat A=\widehat A|_{\cal M}+\widehat A|_{\cal Z}$ 
describes a flat connection on ${\cal M}$ and a pair of oscillators on 
${\cal Z}\times {\cal Y}$ deformed by local as well as topological degrees 
of freedom contained in $\widehat B\,$.
The latter can acquire anti-de Sitter (AdS$_3$) vacuum expectation values, 
\emph{viz.} $\langle \widehat B\rangle=\nu\,$.
Truncating $\widehat B=\nu$ yields Chern--Simons-like HSGRAs on ${\cal M}$ of the 
type introduced originally in \cite{Blencowe:1988gj,Bergshoeff:1989ns,Vasiliev:1989re}.
In the simplest set up, the flat master connection one-form 
\begin{equation}\label{W}
A := \widehat A|_{\cal M}=\tfrac{1}{4i}\sum_{n\geqslant 0;s,t=0,1}
A_{s,t}^{\,\alpha_1 \cdots \alpha_n}  \, \Gamma^s \, k^t  \; q_{(\alpha_1 }
\cdots q_{\alpha_n)}\,\in \, Aq(2;\nu) \, \otimes \, {\cal{C}}\!{\ell}_1\ ,
\end{equation}
takes its values in 
the direct product of the enveloping algebra $Aq(2,\nu)$ of the Wigner-deformed canonical 
coordinates $q_\a$ ($\a=1,2$) 
and Kleinian $k$ \cite{Wigner:50,Vasiliev:1989re,Plyushchay:1997ty}, \emph{viz.}
 \begin{equation}\label{dha}
[q_\alpha, q_\beta ]=2i \epsilon_{\alpha \beta} (1+\nu k) \, ,  \quad
\{q_\alpha, k \} = 0\,, \quad k^2=1 \,,
\end{equation}
and the Clifford algebra ${\cal{C}}\!{\ell}_1\cong \mathbb Z_2$ generated by a single bosonic element $\Gamma$ obeying $\Gamma^2=1\,$.
The gauge algebra contains several inequivalent AdS$_3$ sub-algebras.
The standard choice
\begin{equation}\label{sdrep} 
M^{({\rm St})}_{\a\b}:=q_{(\a}q_{\b)}\ ,
\quad P^{({\rm St})}_{\a\b}:=\Gamma q_{(\a}q_{\b)}\ ,
\end{equation}
yields tensor-spinorial HSGRAs  of Chern--Simons type whose CFT duals \cite{Gaberdiel:2012uj} 
consist of quantum states with bose or fermi statistics, just as in $3+1$ dimensions 
\cite{Vasiliev:1999ba}.
On the other hand, the non-standard choice
\begin{equation}\label{alrep} 
M^{({\rm non-St})}_{\a\b}:=\Pi_+\, q_{(\a}q_{\b)}\ ,
\quad 
P^{({\rm non-St})}_{\a\b}:=\Gamma M^{({\rm non-St})}_{\a\b}\ ,\quad 
\Pi_\pm:=\tfrac12(1\pm k)\ ,
\end{equation}
yields fractional-spin HSGRAs consisting of tensor-spinorial 
fields in $W:=\Pi_+\,A\,\Pi_+$ and Lorentz-singlet 
color gauge fields in $U:=\Pi_-\,A\,\Pi_-$ coupled to the bi-fundamental 
master fields $\psi:=\Pi_+\,A\,\Pi_-$ and $\bar\psi:=\Pi_-\,A\,\Pi_+$ 
which consist of fractional-spin fields in infinite-dimensional 
discrete-series representations of the Lorentz algebra, 
that can be either bosonic or fermionic \cite{wipFS1}.
Assembling the master fields into
\begin{equation}\label{mathbbA}
\mathbb{A}=\left[
\begin{array}{cc}
W & \psi \\
\overline{\psi} & U
\end{array}
\right] \,,
\end{equation}
it follows that ${\rm d}A+A^2=0$ is equivalent to 
${\rm d}\mathbb{A}+ \mathbb{A}^2 = 0 $, \emph{i.e.}
\begin{eqnarray}
&& {\rm d}W+ W \, W +\psi  \, \overline{\psi}=0  \,, \qquad {\rm d}U+U \, U + \overline{\psi}  \, \psi  =0  \, , \label{eq1} \\[5pt]
&& {\rm d}\psi + W \,\psi  + \psi \,U =0  \, , \qquad {\rm d}\overline{\psi} + \overline{\psi}  \,W  + U \, \overline{\psi}=0  \,.\label{eq2}
\end{eqnarray}
As for off-shell formulations, suitable bilinear forms are needed, 
such as the supertrace operation on $Aq(2,\nu)$ \cite{Vasiliev:1989re} 
used to $\nu$-deform Blencowe's tensor-spinorial HSGRA theory \cite{Blencowe:1988gj}.
To analyze the theory on-shell, it suffices, however, to use a representation of 
$Aq(2,\nu)$ in a standard Fock space 
$\mathcal{F}=\bigoplus_{n\geqslant 0}\mathbb{C}\otimes |n\rangle$ 
and its dual $\mathcal{F}^\ast=\bigoplus_{n\geqslant 0}\mathbb{C}\otimes \langle n |$  
\cite{Plyushchay:1997ty}, \emph{viz.}
\begin{equation}\label{a+a-}
  {a}^\pm= \tfrac12(q_1\mp i q_2)   \;,\quad
   {a}^\pm|_{\mathcal F}=\sum_{n\geqslant 0} \sqrt{[n+1]_\nu} \;
   |n+\tfrac12(1\pm 1)\rangle \langle n+\tfrac12(1\mp 1) | ,
 \end{equation}
\begin{equation}k|_{\mathcal F}=\sum_{n\geq 0} (-1)^n |n\rangle \langle n | \, ,\quad [n]_\nu := n+\tfrac{1}{2}(1-(-1)^n)\nu\ ,\quad \langle n | n' \rangle= \delta_{n,n'}\ .
\end{equation}
The master connection \eqref{W} is thus represented by
\begin{equation}\label{Wketbra}
A|_{\mathcal F}=\tfrac{1}{4i} \sum_{m,n\geqslant 0; s=0,1}  A_{s}^{m,n } \, \Gamma^s
 \, |m\rangle \langle
 n|\,, \qquad  A_{s}^{m,n }:=\sum_{p \geqslant 0; t=0,1}  A_{s,t}^{\alpha_1 \cdots \alpha_p}  \, (\mathcal{Q}_{\alpha_1 \cdots \alpha_p}{}^{t})^{m,n }\,.
\end{equation}
where the matrix 
$(\mathcal{Q}_{\alpha_1 \cdots \alpha_p}{}^{t})^{m,n }:= (-1)^{t\,m} \langle
 m |q_{(\alpha_1}\cdots q_{\alpha_p)}  |n \rangle$
represents $Aq(2,\nu)$ in $\mathcal{F}\,$.
Alternatively, using the notation of \eqref{mathbbA}, 
the connection $\mathbb{A}|_{\mathcal F}$ takes the form
\begin{equation}\label{}
\left.\left[
\begin{array}{cc}
W & \psi \\
\overline{\psi} & U
\end{array}
\right]\right|_{\mathcal F} =\tfrac{1}{4i} \sum_{m,n\geqslant 0,  s=0,1} \, \Gamma^s \left[
\begin{array}{cc}
 W_s^{2m,2n}
 \, |2m\rangle \langle
2n | & \psi^{2m, 2n+1 }_s \,  |2m\rangle \langle
2n+1 | \\
\overline{\psi}^{2m+1, 2n}_s  \, |2m+1\rangle \langle
2n|  & U^{2m+1, 2n+1 }_s  \, |2m+1\rangle \langle
2n+1 |
\end{array}
\right] \,.
\end{equation}
The master fields $W$ and $U$ are thus represented faithfully in the even and odd Fock spaces 
$\Pi_+ \mathcal{F}$ and $\Pi_- \mathcal{F}$, respectively.
In particular, the Lorentz connection in $W$ acts non-trivially in $\Pi_+\mathcal F\,$ 
that forms an irrep of the Lorentz-algebra for generic $\nu$, 
and acts trivially in $\Pi_- \mathcal{F}$.
The $(\psi,\bar\psi)$ fields exchange $\Pi_\pm \mathcal F$ and transform under 
one-sided higher-spin and internal gauge transformations.
To compute the Lorentz spin of these intertwiners, one may first use Dirac matrices obeying 
$ \{\gamma_a , \gamma_b\}=-2\eta_{ab}\,$,
$ \eta_{ab}={\rm diag}(-1,+1,+1)$, to convert the standard embedding 
\eqref{sdrep} of the Lorentz algebra into
\begin{equation}\label{sdrepbis}
J^{(\rm{St})}_a:=\tfrac{i}{8} \epsilon^{\alpha \alpha'} (\gamma_a)_{\alpha'}{}^\beta  
M^{({\rm St})}_{\alpha \beta}
=\left( \tfrac{ \{a^+,a^-\} }{4}, \; \tfrac{ a^{+2}+a^{-2} }{2} \,, \; \tfrac{ a^{+2}-a^{-2}}{2i}\right)\ ,
\end{equation}
which acts non-trivially on $\Pi_\pm\mathcal{F}\,$.
For the alternative embedding \eqref{alrep} of the Lorentz algebra, one then has
\begin{equation}
J^{({\rm non-St})}_0|_{\mathcal F}= \sum_{n\geqslant 0} (n+\tfrac{1}{4} (1+\nu)) |2n\rangle \langle 2n |
\;,\end{equation}
\begin{equation}J^{({\rm non-St})}_\pm|_{\mathcal F}= \sum_{n\geqslant 0} 
\sqrt{[2n+2]_\nu[2n+1]_\nu} \; | 2n+1\pm 1\rangle \langle  2n+1\mp 1|\;,
\end{equation}
where $J^{({\rm non-St})}_{\pm} := J^{({\rm non-St})}_1\pm {\rm i} J^{({\rm non-St})}_2
=\tfrac{1}{2}\,\Pi_+(a^{\pm})^2\Pi_+\,$, 
which indeed acts non-trivially on $\Pi_+ \mathcal{F}$ while leaving $\Pi_- \mathcal{F}$ invariant.
Thus, as for the quadratic Casimir operator, one has
\begin{equation}
C_2:=J^a J_{a}\ \Rightarrow\ C_2\psi|_{\mathcal F} = -s_\psi(s_\psi-1)\psi|_{\mathcal F} \,,\qquad s_\psi=\frac{1}{4}(1+\nu) \,,
\end{equation}
that is, $\psi$ has Lorentz spin $\tfrac14(1+\nu)$ where $\nu$ can be any real number.
For negative integer and negative half-integer lowest weights $s_\psi$, singular vectors arise in $\mathcal{F}$, \emph{viz.}
\begin{equation}
a^-|2\ell+1 \rangle = 0 = a^+|2\ell \rangle \ ,\quad \nu=-2\ell - 1\ ,\quad \ell= 0,1,2,...\ .
\end{equation}
Thus, for these values of $\nu$, referred to as critical values, the representation of $Aq(2,\nu)$ in $\mathcal F$ decomposes into a finite-dimensional and an infinite-dimensional algebra as follows \cite{Vasiliev:1989re,Plyushchay:1997ty}:
\begin{equation}\label{criticalAq}
Aq(2;-2\ell-1)|_{\mathcal F}\cong gl(2\ell+1)\oplus Aq(2;2\ell+1)|_{{\mathcal F}}\ .\end{equation}
Thus, using Feigin's notation $gl(\lambda)$ \cite{Feigin}, it follows that
\begin{equation} W|_{\mathcal F}\in gl(-2s_\psi+1;\tau)\otimes {\cal{C}}\!{\ell}_1\ ,\quad U|_{\mathcal F}\in gl(-2s_\psi;\tau)\otimes {\cal{C}}\!{\ell}_1\ ,
\end{equation}
which are infinite-dimensional algebras for generic $s_\psi$ with critical limits given by 
semi-direct sums of a finite-dimensional and an infinite-dimensional sub-algebra with ideal structure 
controlled by $\tau$, which can thus take three distinct values depending on whether one or the other or 
both of the sub-algebras are ideal; in the case at hand, $\tau$ is chosen in accordance 
with \eqref{criticalAq}, that is, such that both sub-algebras are ideals; for further details, 
see \cite{wipFS1}.
Thus, the fractional-spin HSGRA model reduces to a certain conventional tensor-spinorial 
HSGRA in critical limits.

The representation of $f\in Aq(2,\nu)$ in $\mathcal F$ twists the hermitian conjugation 
operation \cite{wipFS1}, \emph{viz.} 
$(f^\dagger)|_{\mathcal F}\equiv ((CfC)|_{\mathcal{F}})^\dagger$ where $(q_\a)^\dagger:=q_\a\,$, 
$k^\dagger:=k$ and $(|m\rangle)^\dagger:=\langle m|$ and the conjugation matrix obeys 
$C^2=1$ and $Ck=kC$ \cite{Plyushchay:1997ty}.
Taking $C^\dagger=C$, we can impose the reality condition
\begin{equation}
(\mathbb{A}|_{\mathcal F})^\dagger= - (\mathbb{C} \mathbb{A} \mathbb{C})|_{\mathcal F} \ ,\quad \mathbb{C}=\left[\begin{array}{cc} \Pi_+\, C& 0\\0&\Pi_-\end{array}\right]\ ,
\end{equation}
for which $U$ is valued in a compact real form of $\Pi_- Aq(2,\nu) \Pi_-$ represented unitarily in 
$\Pi_- \mathcal F$ for all $\nu$, while $W$ is valued in a non-compact real form of 
$\Pi_+ Aq(2,\nu) \Pi_+$ represented unitarily in $\Pi_+ \mathcal F$ iff $\Pi_+ C=\Pi_+$, that is, 
iff $\nu\geqslant -1\,$. 
For $\nu<-1$ there are negative eigenvalues in $\Pi_+ C|_{\mathcal F}$ whose number grows linearly 
with $|\nu|$. Using the notation of analytically continued real forms 
\cite{Fradkin:1990qk,Feigin,wipFS1}, 
one has 
\begin{equation} W|_{\mathcal F}\in u(p_+,p_-;\tau) \otimes  {\cal{C}}\!{\ell}_1 \ ,\quad U|_{\mathcal F}\in u(-2s_\psi;\tau)\otimes  {\cal{C}}\!{\ell}_1\ ,
\end{equation}
\begin{equation} p_+ + p_- = -2s_\psi+1\ ,\quad p_+-p_-=1+(-1)^{-2s_\psi}\ .
\end{equation}
For critical $\nu=-2\ell-1$, one has 
$\Pi_+ C|_{\mathcal F}=\sum_{n=0}^\ell (-1)^n |2n \rangle \langle 2n|  +\sum_{n\geqslant \ell+1}|2n \rangle \langle 2n|$ 
and hence
\begin{equation}
W|_{\mathcal F}\cap gl(2\ell+1) \in u(p_+,p_-)\otimes \mathbb{Z}_2 \ ,\qquad
U|_{\mathcal F}\cap gl(2\ell+1) \in u(\ell)\otimes \mathbb{Z}_2\ ,\end{equation}
where $p_\pm=\tfrac{1}{2}(\ell+1 \pm \tfrac{1+(-1)^\ell}{2})$, and 
$(\psi,\overline{\psi})|_{\mathcal F}\cap gl(2\ell+1)$ belong to bi-fundamental representations 
with integer or half-integer spins and finite-rank color indices.
We note that if $\psi= | \sigma \rangle\langle c |$, where thus $\sigma$ 
is a spin and $c$ is a color, then 
$\bar \psi = - | c \rangle\langle \sigma | C$ and hence 
$\psi \bar \psi = |\sigma \rangle\langle \sigma | C$ while 
$\bar \psi \psi = |\sigma\rangle\langle c|C|c\rangle\langle \sigma|$ that 
can vanish in the non-unitary regime.
Thus, the fractional-spin fields necessarily source the tensor-spinorial fields $W$ 
(\emph{c.f.} positivity of energy in ordinary gravity) 
while the internal gauge field $U$ can be truncated consistently leading to
\begin{equation}
{\rm d}W+W^2 +\psi\bar\psi=0\ ,\quad {\rm d}\psi +W\psi=0\ ,\quad {\rm d}\bar\psi+\bar\psi W=0\ ,\quad \bar\psi\psi=0\ ,
\end{equation} 
which defines a quasi-free differential algebra.

In summary, the model presented here may be of interest in the context of holography where 
$(\psi,\bar\psi)$ are expected to correspond to vertex operators with fractional conformal weights 
resulting in multi-valued correlation functions forming anyonic representations of the braid group, 
possibly along the lines of \cite{Jatkar:1990re}.
The model also serves as a starting point for incorporating local FS degrees of freedom in three and four 
dimensions;
the latter may correspond holographically to massive anyons in three-dimensional
quantum field theories.
As for their quantization, we propose to use generalized Poisson sigma models \cite{wipFS1}.
Subjected to suitable boundary conditions, these sigma models may remain weakly coupled in the critical 
limits of the Prokushkin--Vasiliev system, 
which are otherwise strongly coupled limits of the standard Chern--Simons formulation.

\subsection*{Acknowledgments}

We thank F. Buisseret and D. Jatkar for discussions on various aspects of anyons and fractional-spin 
particles. We thank E. Skvortsov for kindly sending us the paper \cite{Feigin}.
The work of N. B. was supported in part by an ARC contract No. AUWB-2010-10/15-UMONS-1. 
M. V. is supported by FONDECYT postdoctoral grant N 3120103.

\providecommand{\href}[2]{#2}\begingroup\raggedright\endgroup


\begin{thebibliography}{10}

\bibitem{Leinaas:1977fm}
J.~Leinaas and J.~Myrheim, ``{On the theory of identical particles},''
\href{http://dx.doi.org/10.1007/BF02727953}{{\em Nuovo Cim.} {\bf B37} (1977)
  1--23}.

\bibitem{Forte:1990hd}
S.~Forte, ``{Quantum mechanics and field theory with fractional spin and
  statistics},''
\href{http://dx.doi.org/10.1103/RevModPhys.64.193}{{\em Rev.Mod.Phys.} {\bf 64}
  (1992)  193--236}.

\bibitem{Wilczek:1982wy}
F.~Wilczek, ``{Quantum Mechanics of Fractional Spin Particles},''
\href{http://dx.doi.org/10.1103/PhysRevLett.49.957}{{\em Phys.Rev.Lett.} {\bf
  49} (1982)  957}.

\bibitem{Polyakov:1988md}
A.~M. Polyakov, ``{Fermi-Bose Transmutations Induced by Gauge Fields},''
\href{http://dx.doi.org/10.1142/S0217732388000398}{{\em Mod.Phys.Lett.} {\bf
  A3} (1988)  325}.

\bibitem{Semenoff:1988jr}
G.~W. Semenoff, ``{Canonical Quantum Field Theory with Exotic Statistics},''
\href{http://dx.doi.org/10.1103/PhysRevLett.61.517}{{\em Phys.Rev.Lett.} {\bf
  61} (1988)  517}.

\bibitem{Jatkar:1990re}
D.~P. Jatkar and S.~Rao, ``{Anyons and Gaussian conformal field theories},''
\href{http://dx.doi.org/10.1142/S0217732391000257}{{\em Mod.Phys.Lett.} {\bf
  A6} (1991)  289--294}.

\bibitem{Prokushkin:1998bq}
S.~F. Prokushkin and M.~A. Vasiliev, ``{Higher-spin gauge interactions for
  massive matter fields in 3D AdS space-time},''
  \href{http://dx.doi.org/10.1016/S0550-3213(98)00839-6}{{\em Nucl. Phys.} {\bf
  B545} (1999)  385},
\href{http://arxiv.org/abs/hep-th/9806236}{{\tt arXiv:hep-th/9806236}}.

\bibitem{wipFS1}
N.~Boulanger, P.~Sundell, and M.~Valenzuela, ``Three-dimensional
  fractional-spin gravity,'' {\em work in progress}  .

\bibitem{Blencowe:1988gj}
M.~Blencowe, ``{A Consistent Interacting Massless Higher Spin Field Theory in
  $D$ = (2+1)},''
\href{http://dx.doi.org/10.1088/0264-9381/6/4/005}{{\em Class.Quant.Grav.} {\bf
  6} (1989)  443}.

\bibitem{Bergshoeff:1989ns}
E.~Bergshoeff, M.~Blencowe, and K.~Stelle, ``Area preserving diffeomorphisms
  and higher spin algebra,''
\href{http://dx.doi.org/10.1007/BF02108779}{{\em Commun.Math.Phys.} {\bf 128}
  (1990)  213}.

\bibitem{Vasiliev:1989re}
M.~A. Vasiliev, ``Higher spin algebras and quantization on the sphere and
  hyperboloid,'' \href{http://dx.doi.org/10.1142/S0217751X91000605}{{\em
  Int.J.Mod.Phys.} {\bf A6} (1991)  1115--1135}.

\bibitem{Wigner:50}
E.~P. Wigner, ``{Do the Equations of Motion Determine the Quantum Mechanical
  Commutation Relations?},''
  \href{http://dx.doi.org/10.1103/PhysRev.77.711}{{\em Phys. Rev.} {\bf 77}
  (1950)  711--712}.

\bibitem{Plyushchay:1997ty}
M.~S. Plyushchay, ``{Deformed Heisenberg algebra with reflection},''
  \href{http://dx.doi.org/10.1016/S0550-3213(97)00065-5}{{\em Nucl.Phys.} {\bf
  B491} (1997)  619--634},
\href{http://arxiv.org/abs/hep-th/9701091}{{\tt arXiv:hep-th/9701091
  [hep-th]}}.

\bibitem{Gaberdiel:2012uj}
M.~R. Gaberdiel and R.~Gopakumar, ``{Minimal Model Holography},''
  \href{http://dx.doi.org/10.1088/1751-8113/46/21/214002}{{\em J.Phys.} {\bf
  A46} (2013)  214002},
\href{http://arxiv.org/abs/1207.6697}{{\tt arXiv:1207.6697 [hep-th]}}.

\bibitem{Vasiliev:1999ba}
M.~A. Vasiliev, ``{Higher spin gauge theories: Star-product and AdS space},''
\href{http://arxiv.org/abs/hep-th/9910096}{{\tt arXiv:hep-th/9910096}}.

\bibitem{Feigin}
B.~Feigin, ``{The Lie algebras $\frak{gl}(\lambda)$ and cohomologies of Lie
  algebras of differential operators},'' {\em Russ. Math. Surv.} {\bf 43}
  (1988) no.~2, 169--170.

\bibitem{Fradkin:1990qk}
E.~Fradkin and V.~Y. Linetsky, ``{Supersymmetric Racah basis, family of
  infinite dimensional superalgebras, SU(infinity + 1|infinity) and related 2-D
  models},''
\href{http://dx.doi.org/10.1142/S0217732391000646}{{\em Mod.Phys.Lett.} {\bf
  A6} (1991)  617--633}.

\end{thebibliography}

\end{document}